\documentclass[preprint2]{aastex6}
\usepackage{booktabs}

\begin{document}


\title{Tidal Response of Preliminary Jupiter Model}


\author{Sean M. Wahl\altaffilmark{1} }

\affil{Department of Earth and Planetary Science, University of
California, Berkeley, CA, 94720, USA}

\author{William B. Hubbard\altaffilmark{2}}
\affil{Lunar and Planetary Laboratory, The University of
Arizona, Tucson, AZ 85721, USA}

\and

\author{Burkhard Militzer\altaffilmark{3}}

\affil{Department of Earth and Planetary Science, University of
California, Berkeley, CA, 94720, USA}


\altaffiltext{1}{swahl@berkeley.edu}
\altaffiltext{2}{hubbard@lpl.arizona.edu}
\altaffiltext{3}{militzer@berkeley.edu; also Department of Astronomy, University of California, Berkeley, CA,
94720, USA}

\begin{abstract}

    In anticipation of improved observational data for Jupiter's gravitational field
    from the \textit{Juno} spacecraft, we predict the static tidal response for a
    variety of Jupiter interior models based on \textit{ab initio} computer
    simulations of hydrogen-helium mixtures. We calculate hydrostatic-equilibrium
    gravity terms using the non-perturbative \textit{concentric Maclaurin Spheroid}
    (CMS) method that eliminates lengthy expansions used in the theory of figures.
    Our method captures terms arising from the coupled tidal and rotational
    perturbations, which we find to be important for a rapidly-rotating planet like
    Jupiter. Our predicted static tidal Love number $k_2 = 0.5900$ is $\sim$10\%
    larger than previous estimates. The value is, as expected, highly correlated with
    the zonal harmonic coefficient $J_2$, and is thus nearly constant when plausible
    changes are made to interior structure while holding $J_2$ fixed at the observed
    value. We note that the predicted static  $k_2$ might change due to Jupiter's
    dynamical response to the Galilean moons, and find reasons to argue that the
    change may be detectable, although we do not present here a theory of dynamical
    tides for highly oblate Jovian planets.  An accurate model of Jupiter's tidal
    response will be essential for interpreting \textit{Juno} observations and
    identifying tidal signals from effects of other interior dynamics in Jupiter's
    gravitational field.

\end{abstract}

\keywords{giant planets, tides, Jupiter, interiors}



\section{Introduction} \label{sec:intro}

The \textit{Juno} spacecraft began studying Jupiter at close range following its
orbital insertion in early July 2016. The unique low-periapse polar orbit and precise
Doppler measurements of the spacecraft's acceleration will yield parameters of
Jupiter's external gravitational field to unprecedented precision, approaching a
relative precision of $\sim 10^{-9}$ \citep{kaspi2010}. In addition to providing
important information about the planet's interior mass distribution, the
non-spherical components of Jupiter's gravitational field should exhibit a detectable
signal from tides induced by the planet's closer large moons, possibly superimposed
on signals from mass anomalies induced by large-scale dynamic flows in the planet's
interior \citep{cao2015,kaspi2010,kaspi2013}.

As a benchmark for comparison with expected \textit{Juno} data, \citet{hubbard2016}
constructed static interior models of the present state of Jupiter, using a
barotropic pressure-density $P(\rho)$ equation of state for a near-solar mixture of
hydrogen and helium, determined from \textit{ab intio} molecular dynamics simulations
\citep{militzer2013a,militzer2013b}. In this paper, we extend those models to predict
the static tidal response of Jupiter using the three-dimensional concentric Maclaurin
spheroid (CMS) method \citep{wahl2016}.

The \citet{hubbard2016} preliminary Jupiter model is an axisymmetric, rotating model
with a self-consistent gravitational field, shape and interior density profile. It is
constructed to fit pre-\textit{Juno} data for the degree-two zonal gravitational
harmonic $J_2$ \citep{jacobson2003}. While solutions exist matching pre-\textit{Juno}
data for the degree-four harmonic $J_4$, models using the \textit{ab initio} EOS required unphysical
compositions with densities lower than that expected for the pure H-He mixture. As a result, the
preferred model of \citet{hubbard2016} predicts a $J_4$ with an absolute value above
pre-\textit{Juno} error bars. Preliminary Jupiter models consider the effect of a
helium rain layer where hydrogen and helium become immiscible \citep{stevenson1977a}.
The existence of such a layer has important effects for the interior structure of the
planet, since it inhibits convection and mixing between the molecular exterior and
metallic interior portions of the H-He envelope. This circumstance provides a
physical basis for differences in composition and thermal state between the inner and
outer portions of the planet.  Adjustments of the heavy element content and entropy
of the $P(\rho)$ barotrope allow identification of an interior structure consistent
with both pre-\textit{Juno} observational constraints and the \textit{ab initio}
material simulations. The preferred preliminary model predicts a dense inner core
with $\sim$12 Earth masses and an inner hydrogen-helium rich envelope with
$\sim$3$\times$ solar metallicity, using an outer envelope composition matching that
measured by the \textit{Galileo} entry probe.

Although the \textit{Cassini} Saturn orbiter was not designed for direct measurements
of the high degree and order components of Saturn's gravitational field, the first observational
determination of Saturn's second degree Love number $k_2$ was recently reported by
\citet{lainey2016}. This study used an astrometric dataset for Saturn's co-orbital
satellites to fit $k_2$, and identified a value significantly larger than the
theoretical prediction of \citet{gavrilov1977}. The non-perturbative CMS method
obtains values of $k_2$ within the observational error bars for simple models of
Saturn's interior, indicating the high value can be explained completely in terms of
static tidal response \citep{wahl2016}. The perturbative method of
\citet{gavrilov1977} provides an initial estimate of tidally induced terms in the
gravitational potential, but neglects terms on the order of the product of tidal
and rotational perturbations. \citet{wahl2016} demonstrated, that for the
rapidly-rotating Saturn, these terms are significant and sufficient to explain the
observed enhancement of $k_2$.

\section{Barotrope}

We assume the liquid planet is in hydrostatic equilibrium,
\begin{equation} \nabla P = \rho \nabla U,     
    \label{eq:hydrostatic} \end{equation}
where $P$ is the pressure, $\rho$ is the mass density and $U$ the total effective
potential. Modeling the gravitational field of such a body requires a barotrope
$P(\rho)$ for the body's interior. In this paper, we use the barotrope of
\citet{hubbard2016}, constructed from \textit{ab initio} simulations of
hydrogen-helium mixtures \citep{militzer2013a,militzer2013b}. The $P(\rho)$ relation
is interpolated from a grid of adiabats determined from density functional 
molecular dynamics (DFT-MD) simulations using the Perdew-Burke-Ernzerhof (PBE) functional
\citep{PBE} in combination with a thermodynamic integration technique. The
simulations were performed with cells containing $N_{He}=18$ helium and $N_{H}=220$
hydrogen atoms, corresponding to a solar-like helium mass fraction $Y_0=0.245$. An
adiabat is characterized by an entropy per electron $S/k_B/N_e$
\citep{militzer2013b}, where $k_B$ is Boltzmann's constant and $N_e$ is the number of
electrons. Hereafter we refer to this quantity simply as $S$.

\begin{figure}[h!]  
  \centering
    \includegraphics[width=16pc]{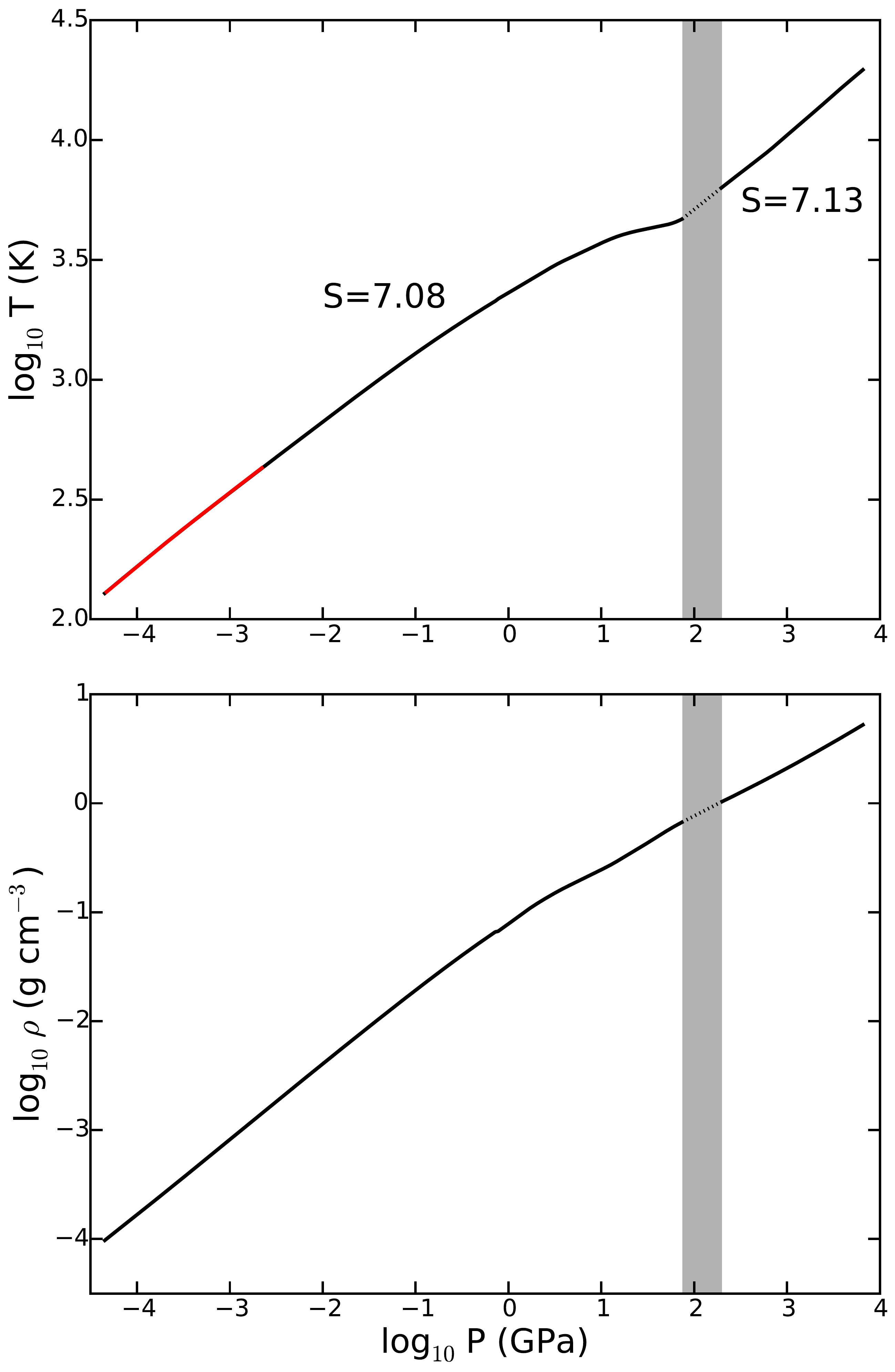}
\caption{ The barotrope used in preferred model Jupiter `DFT-MD\_7.13'. Top:
    temperature-pressure relationship for a hydrogen-Helium mixture with Y=0.245,
    with a entropy $S=7.08$ at pressures below the demixing region, and $S=7.13$ at
    pressures above the demixing region. The helium demixing region is shown by the
    gap and shaded region. The red line shows measurements from the \textit{Galileo}
probe. Bottom: density-pressure relationship for the same barotrope.}
\label{fig:eos}
\end{figure}

In our treatment, the term ``entropy'' and the symbol $S$ refer to a particular
adiabatic temperature $T(P)$ relationship for a fixed composition H-He mixture
($Y_0=0.245$) as determined from the \textit{ab initio} simulations. The value of $S$
in the outer portion of the planet is determined by matching the $T(P)$ measurements
from the Galileo atmospheric probe (see Figure \ref{fig:eos}). This adiabatic $T(P)$
is assumed to apply to small perturbations of composition, in terms of both helium
fraction and metallicity. \citet{hubbard2016} demonstrated that these compositional
perturbations have a negligible effect on the temperature distribution in the
interior.

The density perturbations to the equation of state are estimated using the additive
volume law,
\begin{equation} 
    V(P, T) = V_H(P, T) + V_{He}(P, T) + V_{Z}(P,T),
 \label{eq:volume_law}
\end{equation}
where the total volume $V$ is the sum of partial volumes of the main components $V_H$
and $V_{He}$, the heavy element component $V_Z$. \citet{hubbard2016} demonstrated
that this leads to a modified density $\rho$ in terms of the original H-He EOS
density $\rho_0$,
\begin{equation} 
    \frac{\rho_0}{\rho} = \frac{1-Y-Z}{1-Y_0} + 
    \frac{ZY_0 + Y - Y_0}{1-Y_0}\frac{\rho_0}{\rho_{He}} + Z\frac{\rho_0}{\rho_Z},
 \label{eq:density_ratio}
\end{equation}
in which all densities are are evaluated at the same $T(P)$ and $Y_0$ is the helium
fraction used to calculate the H-He equation of state.

The choice of equation of state effects the density structure of the planet, and
consequently, the distribution of heavy elements that is consistent with observational
constraints. For comparison, we also construct models using the \citet{saumon1995}
equation of state (SCvH) for H-He mixtures, which has been used extensively in giant planet
modeling. 

\textit{Ab initio} simulations show that, at the temperatures relevant to Jupiter's
interior, there is no distinct, first-order phase transition between molecular
(diatomic, insulating) hydrogen to metallic (monatomic, conducting) hydrogen
\citep{vorberger2007}. In the context of a planet-wide model, however, the transition
takes place over the relatively narrow pressure range between $\sim$1-2 Mbar. Within
a similar pressure range an immiscible region opens in the H-He phase diagram
\cite{morales2013}, which under correct conditions allows for a helium rain layer
\cite{stevenson1977a,stevenson1977b}. By comparing our adiabat calculations to the
\cite{morales2013} phase diagram, we predict such a helium rain layer in present-day
Jupiter \citep{hubbard2016}. The extent of this layer in our models is highlighted in
Figure \ref{fig:eos}. While the detailed physics involved with the formation and
growth of a helium rain layer is poorly understood, the existence of a helium rain
layer has a number of important consequences for the large-scale structure of the
planet. In our models, we assume this process introduces a superadiabatic temperature
gradient and a compositional difference between the outer, molecular layer and inner,
metallic layer.

In summary, the barotrope and resulting suite of axisymmetric Jupiter models that we
use in this investigation are identical to the results presented by
\citet{hubbard2016}. Each model has a central core mass and envelope metallicities
set to fit the observed $J_2$ \citep{jacobson2003}, with densities corrected to be
consistent with non-spherical shape of the rotating planet. Since tidal corrections
to a rotating Jupiter model are of order $10^{-7}$, see
Table~\ref{tab:jupiter_params} and the following section, it is unnecessary to re-fit
the tidally-perturbed models to the barotrope assumed for axisymmetric models. 

\begin{figure}[h!]  
  \centering
    \includegraphics[width=19pc]{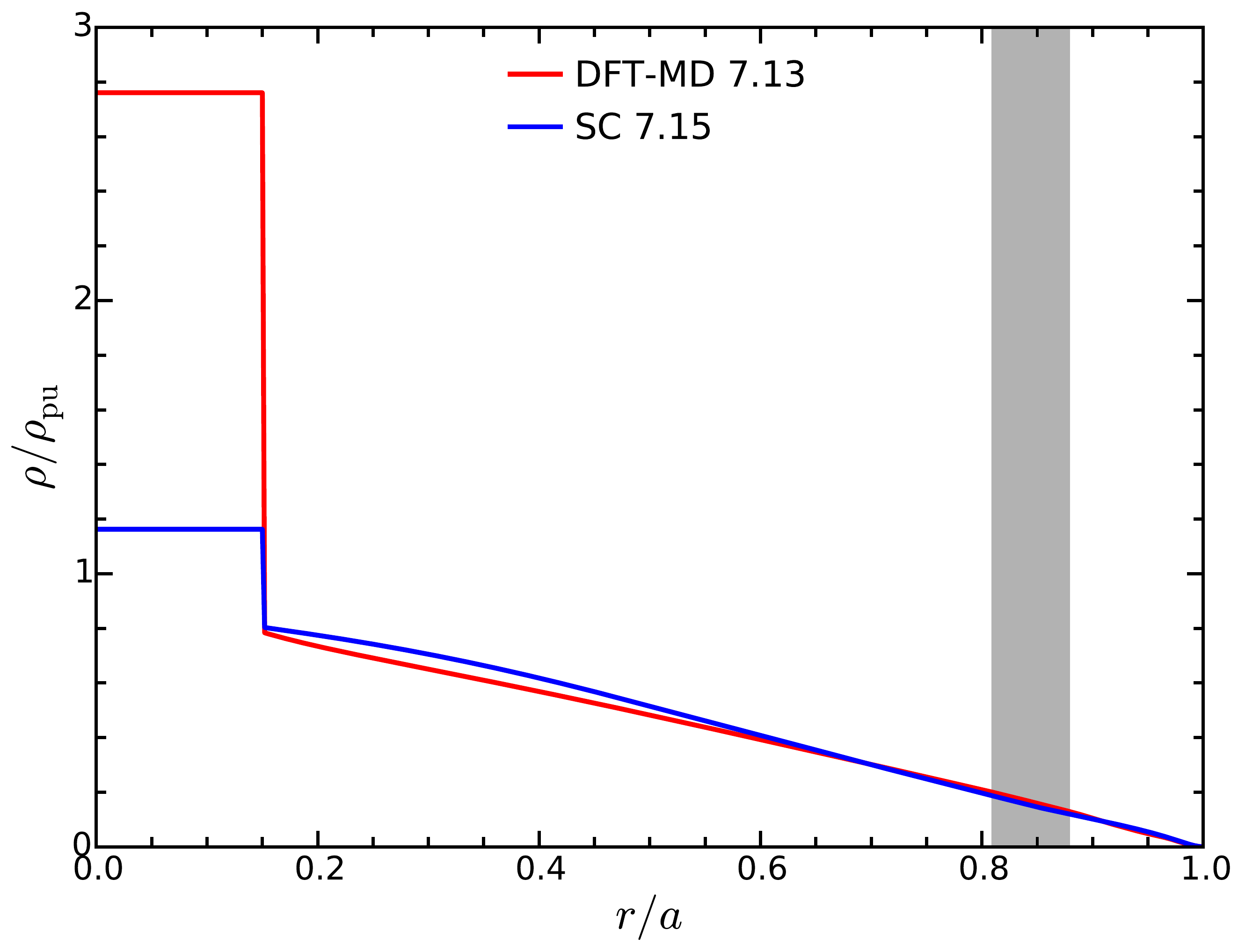}
\caption{   Density structure of Jupiter models (the planetary unit of
density $\rho_{pu}=M/a^3$).  The red curve shows our preferred
    model based on \textit{ab initio} calculations. The blue curve uses the Saumon and
    Chabrier equation of state. The shaded area denotes the helium demixing region.
    Both models have $N=511$ layers and a dense core within $r=0.15a$.  Constant core
    densities are adjusted to match $J_2$ as measured by fits to Jupiter flyby Doppler
    data \citep{jacobson2003}.}
\label{fig:density_structure}
\end{figure}

The physical parameters for each of these models is summarized in Table
\ref{tab:model_values}. The gravitational moments at the planet's surface are
insensitive the precise distribution of extra heavy-element rich material within the
innermost part of the planet. For instance, we cannot discern between dense rocky
core with a and that same heavy material dissolved in metallic hydrogen and spread
over a larger, but still restricted volume. Maintaining a constant core radius is
computationally convenient when finding a converged core mass to $J_{2}$, since it
requires no modification of the radial grid used through the envelope. For this
reason we consider models with a constant core radius of $0.15a$. Decreasing this
radius below $0.15a$ for a given core mass has a negligible effect on the calculated
gravitational moments \citep{hubbard2016}. Figure \ref{fig:density_structure} shows
the density profile for two representative models.  In general, models using the
DFT-MD equation of state lead to a larger central core and a lower envelope
metallicity than those using SCvH.  \citet{hubbard2016} also noted that these models
predict a value for $J_4$ outside the reported observational error bars
\citep{jacobson2003}, since they would require unrealistic negative values of $Z$ to
match both $J_2$ and $J_4$.

\section{Static Tide Calculations}

To calculate the gravitational moments, we use the non-perturbative \textit{concentric
Maclaurin spheroid} method which was introduced by \citet{hubbard2012,hubbard2013}
and extended to three dimensions by \citet{wahl2016}. In this method, the density
structure is parameterized by $N$ nested constant-density spheroids. For a given set of
spheroids, the gravitational field is calculated as a volume-integrated function
of all of the spheroids. The method then iterates to find the shape of each spheroid
such that the surface of each is an equipotential surface under the combined effect of
the planet's self-gravity, the centrifugal potential from rotation and the external
gravitational perturbation from a satellite. The result is a model with
self-consistent shape, internal density distribution and gravitational field
described up to a chosen harmonic degree and order limit, $n_{\rm lim}$.

The non-spherical components of the gravitational potential are described by
two non-dimensional numbers
\begin{equation} q_{\rm rot} = \frac{\omega^2 a^3}{GM}, \label{eq:qrot} \end{equation}
describing the relative strength of the rotational perturbation, and
\begin{equation} q_{\rm tid} = -\frac{3m_{\rm s}a^3}{MR^3}, \label{eq:qtid}
\end{equation}
the analogous quantity for the tidal perturbation. Here $G$ is the universal
gravitational constant, $M$ is the total mass of the planet, $a$ is the maximum
equatorial radius, $m_{\rm s}$ is the mass of the satellite and $R$ is the orbital
distance of the satellite. The parameterization is completed by a third
non-dimensional number $R/a$, representing the ratio of satellite distance to
equatorial radius. For non-zero $q_{\rm tid}$, the calculated figure changes from a
axisymmetric about the rotational axis to a fully triaxial spheroid.

\begin{figure}[h!]  
  \centering
    \includegraphics[width=16pc]{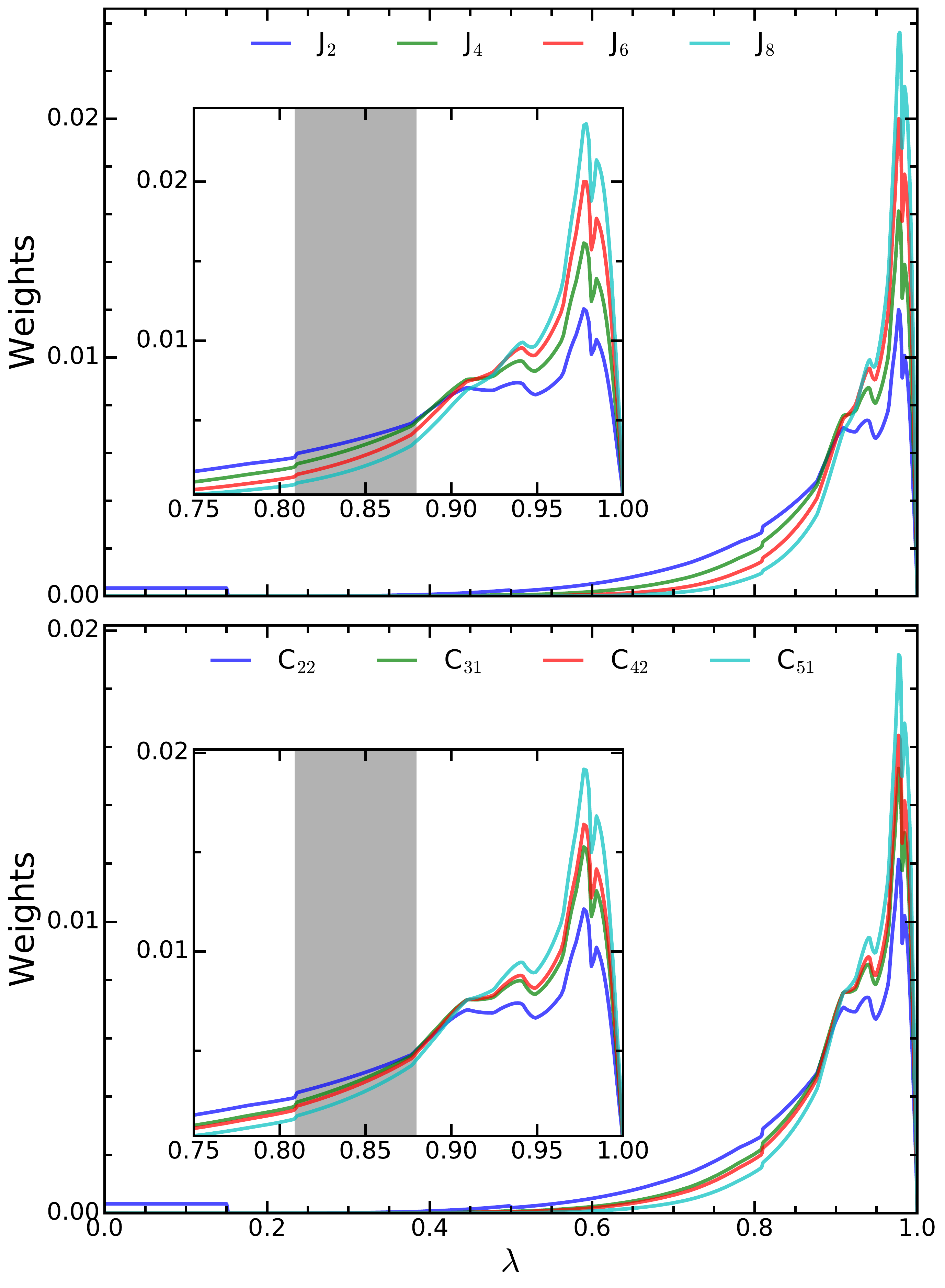}
\caption{Top: Relative contribution of spheroids to external gravitational zonal
    harmonic coefficients up to order 8. Bottom: Relative contribution of spheroids
    to to external gravitational tesseral coefficients up to order 4. Tesseral
    moments of the same order (i.e. $C_{31}$ and $C_{33}$) have indistinguishable
radial distributions. Values normalized so that each harmonic integrates to unity.
The shaded area denotes the helium demixing region.  }
\label{fig:jupiter_weights}
\end{figure}

From our CMS simulations, we find the zonal $J_n$ and tesseral $C_{nm}$ and $S_{nm}$
gravitational harmonics. These harmonics sample slightly different regions of the
planet. Figure~\ref{fig:jupiter_weights} show the relative weight of the contribution to the
low order $J_n$ and $C_{nm}$ as a function of non-dimensional radius.  In the case of
Jupiter and the Galilean satellites, $q_{\rm rot} \gg q_{\rm tid}$, and tidal
perturbations from multiple moons can be linearly superposed.  Moreover, all of the
gravitationally important moons have orbits with nearly zero inclination. This allows
us to treat a simplified case where we consider a single satellite with a fixed
position in the equatorial plane, at angular coordinates $\mu=\cos \theta=0$ (where
$\theta$ is the satellite's colatitude measured from Jupiter's pole), and $\phi=0$
(the satellite's Jupiter-centered longitude).  By symmetry, this configuration
constrains $S_{nm}=0$, and the tidal Love numbers can then be determined from 
\begin{equation}
    k_{nm} = -\frac{2}{3}\frac{(n+m)!}{(n-m)!}\frac{C_{nm}}{P_n^m(0)q_{\rm tid}}
    \left( \frac{a}{R} \right)^{2-n},
\label{eq:kn}
\end{equation}
where $P^m_n(0)$ is the associated legendre polynomial evaluated at $\mu=0$. In this
paper we perform independent calculations for the three satellites with the largest
$q_{\rm tid}$: Io, Europa and Ganymede (see Table~\ref{tab:jupiter_params}).

For a tidally-perturbed non-rotating body, $k_{nm}$ is degenerate with respect to
$m$. However, \citet{wahl2016} found that a large rotational bulge breaks this
degeneracy.  This leads to unexpectedly large values for some of the higher order
$k_{nm}$.  In the case of a rapidly-rotating gas giant, the predicted splitting of
the $k_{nm}$ and shift of $k_{22}$ is well above the expected uncertainty of
\textit{Juno} measurements.

\section{Results}

\subsection{State Mixing for Static Love Numbers} \label{state_mixing}

\begin{figure}[h!]  
  \centering
    \includegraphics[width=19pc]{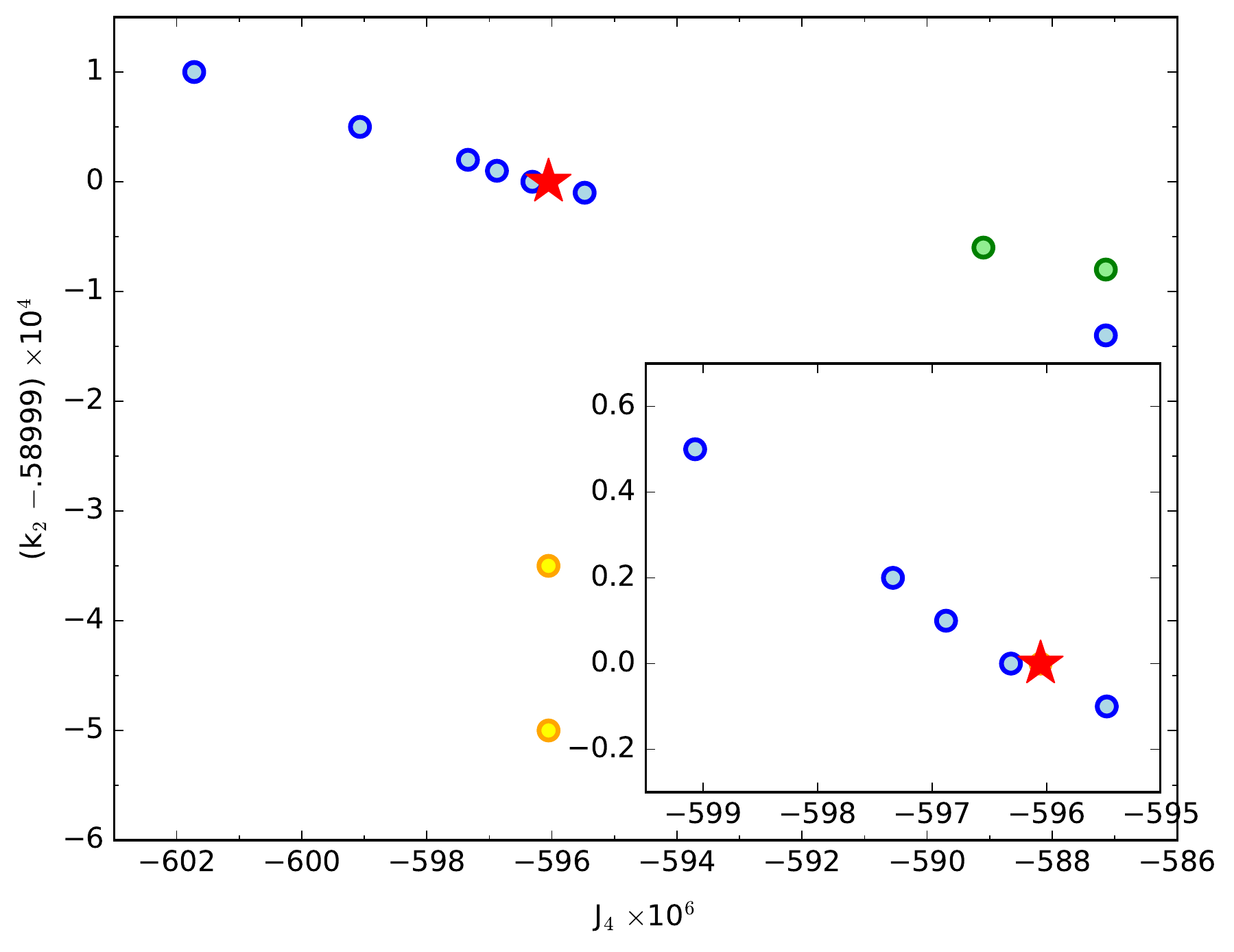}
\caption{ Predicted $k_2$ Love numbers for Jupiter models plotted against $J_4$. The
    favored interior model `DFT-MD\_7.13' with a tidal perturbation from Io is
    denoted by the red star. The other interior models with barotropes based on the
    DFT-MD simulations (blue) have $k_2$ forming a linear trend with $J_4$.  Models
    using the Saumon and Chabrier barotrope (green) plot slightly above this trend.
    The of $k_2$ for a single model `DFT-MD\_7.13' with tidal perturbations from
    Europa and Ganymede (yellow) show larger differences than any resulting from
    interior structure.
    \label{fig:J4_k2}}
\end{figure}

In the CMS method applied to tides, we calculate the tesseral harmonics $C_{nm}$
directly, and the Love numbers $k_{nm}$ are then calculated using Eq.~\ref{eq:kn}.
For the common tidal problem where $q_{\rm tid}$ and $q_{\rm rot}$ are carried to
first order perturbation only, this definition of $k_{nm}$ removes all dependence on
the small parameters $q_{\rm tid}$ and $a/R$, which is convenient for calculating the
expected tidal tesseral terms excited by satellites of arbitrary masses at arbitrary
orbital distances.  However, the high-precision numerical results from our CMS tidal
theory reveal that when $q_{\rm rot} \approx 0.1$, as is the case for Jupiter and
Saturn, the mixed excitation of tidal and rotational harmonic terms in the external
gravity potential has the effect of introducing a small but significant dependence of
$k_{22}$ on $a/R$; see Fig.~\ref{fig:J4_k2}. In the absence of rotation, the CMS
calculations yield results without any state mixing, and the $k_{nm}$ are, as
expected, constant with respect to $a/R$.  It is important to note this effect on the
{\it static} Love numbers because, as we discuss below, dynamical tides can also
introduce a dependence on $a/R$ via differing satellite orbital frequencies.

\subsection{Calculated Static Tidal Response}

\begin{figure}[h!]  
  \centering
    \includegraphics[width=19pc]{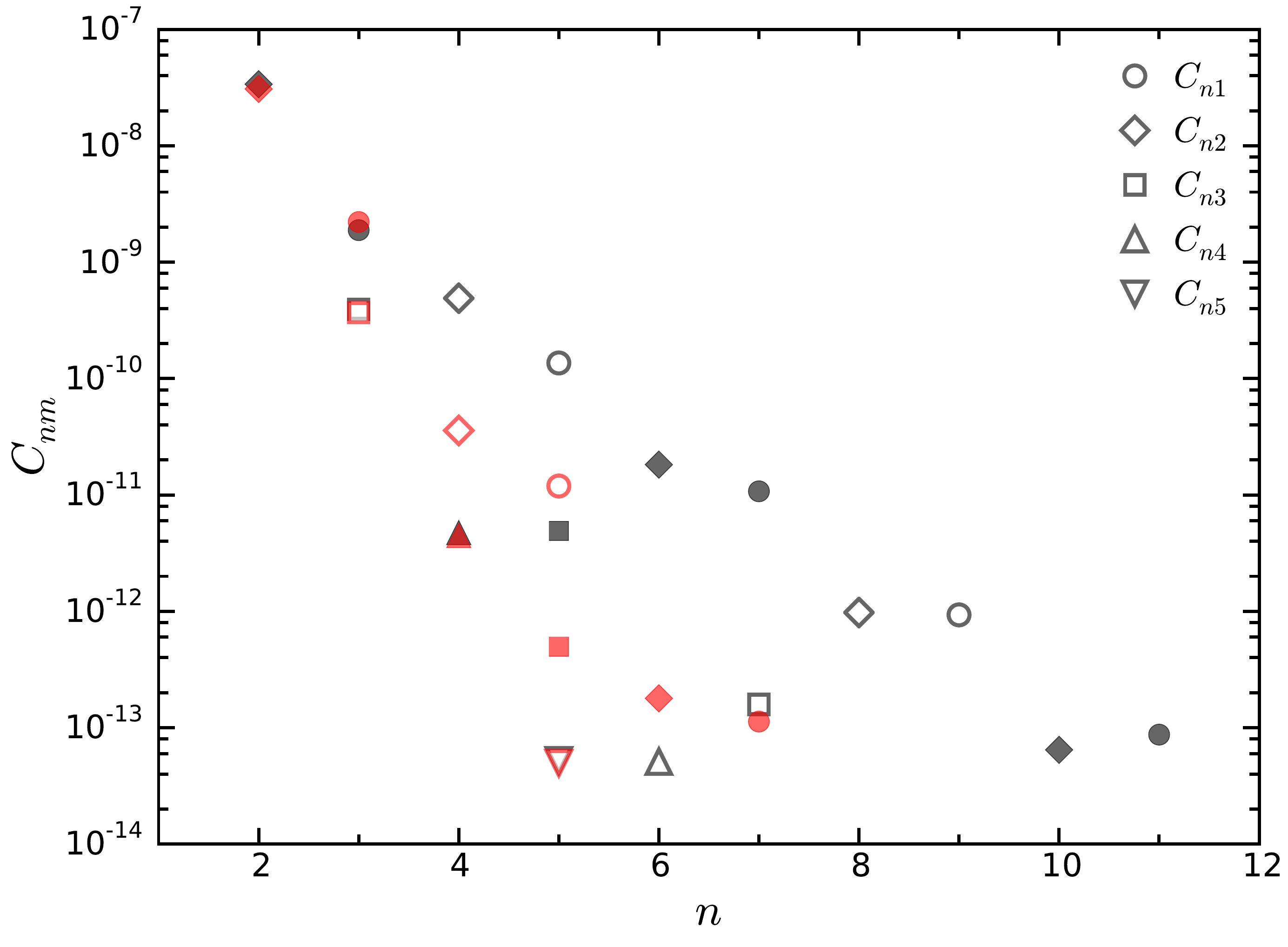}
\caption{ The tesseral harmonic magnitude $C_{nm}$ for the `DFT\_MD 7.13' Jupiter
model with a tidal perturbation corresponding to Io at its average orbital distance.
Black: the values calculated with Jupiter's rotation rate; red: the values for
a non-rotating body with identical layer densities.  Positive values are shown as
filled and negative as empty.}
\label{fig:tesseral_rotation}
\end{figure}

The calculated zonal harmonics $J_n$ and tidal Love numbers $k_{nm}$ for all of the
Jupiter models with Io satellite parameters are shown in
Tab.~\ref{tab:model_harmonics}. Our preferred Jupiter model has a calculated $k_{2}$
of 0.5900. In all cases, these Love numbers are significantly different from those
predicted for a non-rotating planet (see Tab.~\ref{tab:satellite_harmonics}).
Fig.~\ref{fig:tesseral_rotation} shows the different tesseral harmonics $C_{nm}$
calculated with and without rotation. For a non-rotating planet with identical
density distribution to the preferred model we find a much smaller $k_{22}=0.53725$.
\textit{Juno} should, therefore, be able to test for the existence of the rotational
enhancement of the tidal response.

The effect of the interior mass distribution for a suite of realistic models has a
minimal effect on the tidal response. Most models using the DFT-MD barotrope are
within a 0.0001 range of values. The one outlier being the model constrained to match
$J_4$ with unphysical envelope composition. The models using the SCvH barotrope
yields slightly lower, but still likely indistinguishable values of $k_{22}$. The
higher order harmonics show larger relative differences between models, but still
below detection levels. Regardless, the zonal harmonic values are more diagnostic for
differences between interior models than the tidal Love numbers. Fig.~\ref{fig:J4_k2}
summarizes these results, and shows that the calculated $k_{22}$ value varies
approximately linearly with $J_4$.  If \textit{Juno} measures higher order tesseral
components of the field, it may be able to verify a splitting of the $k_{nm}$ Love
numbers with different $m$, for instance, a predicted difference between
$k_{31}\sim0.19$ and $k_{33}\sim0.24$.

\begin{figure}[h!]  
  \centering
    \includegraphics[width=19pc]{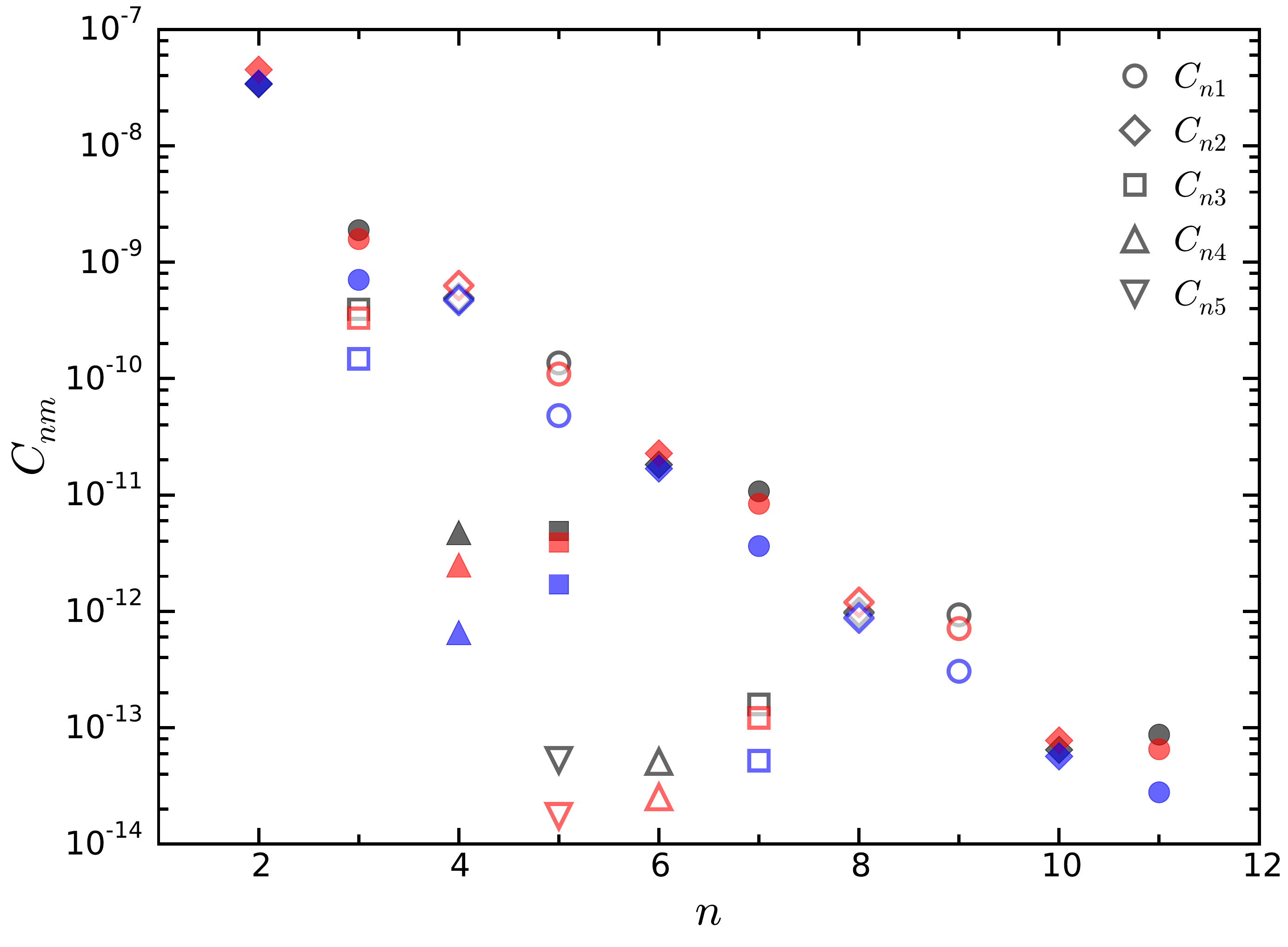}
\caption{ The tesseral harmonic magnitude $C_{nm}$ for the `DFT\_MD 7.13' Jupiter
model with a tidal perturbation corresponding to different satellites: Io (black),
Europa (red) and Ganymede (blue).}
\label{fig:tesseral_satellites}
\end{figure}

In addition, we find small, but significant, differences between the tidal response
between Jupiter's most influential satellites. Fig. \ref{fig:tesseral_satellites}
shows the calculated $C_{nm}$ for simulations with Io, Europa and Ganymede. We
attribute the dependence on orbital distance to the state mixing described in Section
\ref{state_mixing}. This leads to a difference in $k_{22}$ between the three
satellites (Tab.~\ref{tab:satellite_harmonics}) that may be discernible in
\textit{Juno}'s measurements.

\section{Correction for Dynamical Tides}

\subsection{Small Correction for Non-rotating Model of Jupiter}

The general problem of the tidal response of a rotationally-distorted liquid
Jovian planet to a time-varying perturbation from an orbiting satellite
has not been solved to a precision equal to that of the static CMS tidal
theory of \citet{wahl2016} and this paper.  However, an elegant approach
based on free-oscillation theory has been applied to the less general problem
of a non-rotating Jovian planet perturbed by a satellite in a circular orbit
\citep{vorontsov1984}.  Let us continue to use the spherical coordinate system $(r,\theta,\phi)$,
where $r$ is radius, $\theta$ is colatitude and $\phi$ is longitude.
Assume that the satellite is in the planet's equatorial
plane ($\theta=\pi/2$) and orbits prograde at angular rate $\Omega_S$.  For a given planet
interior structure, \citet{vorontsov1984} first obtain its eigenfrequencies
$\omega_{\ell m n}$ and orthonormal
eigenfunctions ${\bf u}_{\ell m n}(r,\theta,\phi)$, projected on spherical harmonics
of degree $\ell$ and order $m$ (the index $n=0,1,2,...$ is the number of radial nodes
of the eigenfunction).  Note that in their convention, oscillations moving prograde
(in the direction of increasing $\phi$) have negative $m$, whereas some authors, e.g.
\citet{marley1993} use the opposite convention.

Treating the tidal response as a forced-oscillation problem, equation (24) of
\citet{vorontsov1984}, the vector tidal displacement ${\bf \xi}$ then reads

\begin{equation}
    {\bf \xi}({\bf r},t) = -\sum_{\ell,m,n} {{({\bf u}_{\ell,m,n},{\bf \nabla}\psi^r_{\ell m})}
\over {\omega^2_{\ell m n} - m^2 \Omega_S^2}} e^{-i m \Omega_S t},
\label{eq:xioft}
\end{equation}
where $({\bf u}_{\ell,m,n},{\bf \nabla} \psi^r_{\ell m})$ is the integrated scalar product of the
vector displacement eigenfunction ${\bf u}_{\ell m n}(r,\theta,\phi)$ and the
gradient of the corresponding term of the satellite's tidal potential
$\psi^r_{\ell m}(r,\theta,\phi,t)$, {\it viz.}
\begin{equation}
    ({\bf u}_{\ell,m,n},{\bf \nabla} \psi^r_{\ell m})=
\int dV \rho_0(r) ({\bf u}_{\ell,m,n} \cdot {\bf \nabla} \psi^r_{\ell m}).
\label{eq:scalprod}
\end{equation}
The integral is taken over the entire spherical volume of the planet,
weighted by the unperturbed spherical mass density distribution $\rho_0(r)$.

\citet{vorontsov1984} then show that, for the nonrotating Jupiter problem, the
degree-two dynamical Love number $k_{2,d}$ is determined to high precision ($\sim$
0.05\%) by off-resonance excitation of the $\ell=2, m=2, n=0$ and $\ell=2, m=-2, n=0$
oscillation modes, such that
\begin{equation}
    k_{2,d}={{\omega^2_{220}} \over {\omega^2_{220} - (2 \Omega_S)^2}} k_2,
\label{eq:k2d}
\end{equation}
noting that $\omega_{220}$ and $\omega_{2-20}$ are equal for nonrotating Jupiter (all
Love numbers in the present paper written without the subscript {\it d} are
understood to be static).  For a Jupiter model fitted to the observed value of $J_2$,
\citet{vorontsov1984} set $\Omega_S = 0$ to obtain $k_2  = 0.541$, within 0.7\% of
our nonrotating value of 0.53725 (see Table~\ref{tab:satellite_harmonics}).  Setting
$\Omega_S$ to the value for Io, Eq.~\ref{eq:k2d} predicts that $k_{2,d} = 0.547$,
i.e. the dynamical correction increases $k_2$ by 1.2\%.  This effect would be only
marginally detectable by the \textit{Juno} measurements of Jupiter's gravity, given
the expected observational uncertainty.

\subsection{Dynamical Effects for Rotating Model of Jupiter}

For a more realistic model of Jupiter tidal interactions, the dynamical correction to
the tidal response might be larger, and therefore, more detectable.  We have already
shown (Table 4) that inclusion of Jupiter's rotational distortion increases the
static $k_2$ by nearly 10\% above the non-rotating static value for a spherical
planet.  In this section, we note that
Jupiter's rapid rotation may also change Jupiter's dynamic tidal response,
by a factor that remains to be calculated.

In a frame co-rotating with Jupiter at the rate $\Omega_P=2 \pi / 35730$s,
the rate at which the subsatellite point moves is obtained by the scalar difference
$\Delta \Omega = \Omega_S - \Omega_P$, which is negative for all Galilean satellites.  Thus,
in Jupiter's fluid-stationary frame, the subsatellite point moves retrograde
(it is carried to the west by Jupiter's spin).  
For Io, we have $\Delta \Omega = -1.35 \times 10^{-4}$ rad/s.
Jupiter's rotation splits the
$\omega_{2\pm20}$ frequencies \citep{vorontsov1981}, such that
$\omega_{2-20}= 5.24 \times 10^{-4}$ rad/s and
$\omega_{220}= 8.73 \times 10^{-4}$ rad/s.  The oscillation
frequencies of the Jovian modes closest to tidal resonance with Io are
higher than the frequency of the tidal disturbance in
the fluid-stationary frame, but are closer to resonance than
in the case of the non-rotating model considered by
\citet{vorontsov1984}.

An analogous investigation for tides on Saturn raised by
Tethys and Dione yields results similar to the Jupiter values:
tides from Tethys or Dione are closer to resonance with normal modes for $\ell=2$ and
$m=2$ and $m=-2$.  Since our static
value of $k_2$ for Saturn \citep{wahl2016} is robust to various assumptions about interior
structure and agrees well
with the value deduced by \citet{lainey2016}, so far we have no evidence for dynamical
tidal amplification effects in the Saturn system.  

Unlike the investigation of \citet{lainey2016}, which relied on analysis of astrometric data for
Saturn satellite motions, the \textit{Juno} gravity investigation will attempt to directly determine
Jupiter's $k_2$ by analyzing the influence of Jovian tesseral-harmonic terms on the spacecraft orbit.
A discrepancy
between the observed $k_2$ and our predicted static $k_2$
would indicate the need for
a quantitative theory of dynamical tides in rapidly rotating Jovian planets. 

\section{Conclusions}

Our study has predicted the static tidal Love numbers $k_{nm}$ for Jupiter and its three
most influential satellites. These results have the following features: (a) They are
consistent with the most recent evaluation of Jupiter's $J_2$ gravitational
coefficient; (b) They are fully consistent with state of the art interior models
\citep{hubbard2016} incorporating DFT-MD equations of state, with a density
enhancement across a region of H-He imiscibility \citep{morales2013}; (c) We use the
non-perturbative CMS method for the first time to calculate high-order tesseral
harmonic coefficients and Love numbers for Jupiter.

The combination of the DFT-MD equation of state and observed $J_{2n}$ strongly limit
the parameter space of pre-\textit{Juno} models. Within this limited parameter space,
the calculated $k_{nm}$ show minimal dependence on details of the interior structure.
Despite this, our CMS calculations predict several interesting features of Jupiter's
tidal response that the \textit{Juno} gravity science system should be able to
detect. In response to the rapid rotation of the planet the $k_2$ tidal Love number
is predicted to be much higher than expected for a non-rotating body. Moreover, the
rotation causes state mixing between different tesseral harmonics, leading to a
dependence of higher order static $k_{nm}$ on both $m$ and the orbital distance of the
satellite. An additional, significant dependence on $a/r$ is expected in the dynamic
tidal response. We present an estimate of the dynamical correction to our
calculations of the static response, but a full analysis of the dynamic theory of
tides has yet to be performed.

\acknowledgments
This work was supported by NASA’s Juno project. Sean Wahl and Burkhard Militzer
acknowledge the support the National Science Foundation (astronomy and astrophysics
research grant 1412646).






\bibliographystyle{aasjournal}                       
\bibliography{jupiter}

\floattable
\begin{deluxetable}{l|rrr}
\tablecaption{Jupiter Model Parameters}
\tablehead{
    \colhead{} & \colhead{Jupiter} & \colhead{} & \colhead{}}
\startdata
$GM$ & $1.26686535 \times 10^{8}$ \tablenotemark{a} &  ${\rm (km^3/s^2) }$  & \\
$a$ & $7.1492 \times 10^{4}$       \tablenotemark{a}  & ${\rm (km) }$   & \\ 
$J_2  \times  10^6$  & $14696.43$ \tablenotemark{a}  &        & \\
$J_4  \times  10^6$  & $-587.14$\tablenotemark{a}  &        & \\
$q_{\rm rot} $  & $ .08917920 $ \tablenotemark{b}  & & \\
$r_{\rm core} / a$ & $0.15$  & \\ 
\tableline
\tableline
& \colhead{Io\tablenotemark{b}} & \colhead{Europa\tablenotemark{b}} & \colhead{Ganymede\tablenotemark{b}} \\
\tableline
$q_{\rm tid} $  & $-6.872 \times 10^{-7} $ & $-9.169\times 10^{-8 }$  
&  $-6.976\times10^{-8}$  \\
$R/a$  & $5.90$  &  $9.39$  & $14.98$ \\
\enddata
\tablerefs{ a. \citet{jacobson2003}, b. \citet{archinal2011}}
\label{tab:jupiter_params}
\end{deluxetable}

\floattable
\begin{deluxetable}{l|cc|rrrr}
\tablecaption{Jupiter Model Values}
\tablehead{ 
    \colhead{} & \colhead{$S_{\rm molec.}$} &  \colhead{$S_{\rm metal.}$} & \colhead{$M_{\rm core}$} & \colhead{$M_{\rm Z,molec.}$}
    & \colhead{$M_{\rm Z,metal.}$} & \colhead{$Z_{\rm global}$} \\
    &  \colhead{($S/k_B/N_e$)} & \colhead{($S/k_B/N_e$)} & \colhead{($M_E$)} &  \colhead{($M_E$)} &  \colhead{($M_E$)} & }
\startdata
DFT-MD   7.24             &  7.08  &  7.24  &  12.5  &  0.9     &  10.3  &  0.07  \\
DFT-MD  7.24~(equal-$Z$)  &  7.08  &  7.24  &  13.1  &  1.1     &  7.5   &  0.07  \\
DFT-MD  7.20              &  7.08  &  7.20  &  12.3  &  0.8     &  9.9   &  0.07  \\
DFT-MD  7.15              &  7.08  &  7.15  &  12.2  &  0.7     &  9.2   &  0.07  \\
DFT-MD  7.15~($J_4$)      &  7.08  &  7.15  &  9.7   &  $-$0.6  &  14.9  &  0.08  \\
{\bf DFT-MD 7.13}       & {\bf 7.08}   & {\bf 7.13} & {\bf 12.2}  & {\bf  0.7}   &
{\bf 8.9}   &  {\bf 0.07}  \\
DFT-MD  7.13~(low-$Z$)  &  7.08  &  7.15  &  14.0  &  0.2  &  1.1   &  0.05  \\
DFT-MD  7.08            &  7.08  &  7.08  &  12.0  &  0.6  &  8.3   &  0.07  \\
SC      7.15            &  7.08  &  7.15  &  4.8   &  3.5  &  28.2  &  0.11  \\
SC      7.15~($J_4$)    &  7.08  &  7.15  &  4.3   &  3.2  &  29.3  &  0.12  \\
\enddata
\tablenotetext{}{Model parameters from\cite{hubbard2016}. $S$ is the
specific entropy for the adiabat through the inner or outer H-He envelope. $M$ is the
mass of heavy elements included in each layer. Each model matches
observed  $J_2 = 14696.43 \times 10^{−6}$ \citep{jacobson2003}, JUP230 orbit solution,
to six significant figures. Models denoted as 'DFT-MD' if equation of state based on
\textit{ab initio} simulations or 'SC' for the \citet{saumon1995} equation of
state, with a number denoting the entropy below the helium demixing layer.  The
number of Models denoted with ($J_4$) also match observed $J_4=-596.31\times
10^{-6}$. Model denoted (equal-$Z$) is constrained to have same metallicity in inner
and outer portions of the planet. Preferred interior model shown in bold face.}
\label{tab:model_values}
\end{deluxetable}

\floattable
\begin{deluxetable}{l|rrrr|rrrrrrrrrrrr}
\rotate
\tablecaption{Gravitational Harmonic Coefficients and Love Numbers}
\tablehead{ 
\colhead{(all $J_n$ $\times$ $10^6$)} &  $J_{4}$ &  $J_{6}$ &  $J_{8}$ &  $J_{10}$ &  $k_{22}$ &  $k_{31}$ &  $k_{33}$ &
$k_{42}$ &  $k_{44}$ &  $k_{51}$ &  $k_{53}$ &  $k_{55}$ &  $k_{62}$ &  $k_{64}$ &
$k_{66}$} 
\startdata
pre-\textit{Juno}~observed &  -587.14 &    34.25 &   \nodata & \nodata & \nodata &
\nodata & \nodata & \nodata & \nodata & \nodata & \nodata & \nodata & \nodata &
\nodata & \nodata  \\
\colhead{(JUP230)\tablenotemark{a}} &  $\pm$1.68 &   $\pm$5.22 &   \nodata & \nodata & \nodata &
\nodata & \nodata & \nodata & \nodata & \nodata & \nodata & \nodata & \nodata &
\nodata & \nodata  \\
DFT-MD 7.24            &  -597.34 &    35.30 &   -2.561 &     0.212 &   0.59001 &   0.19455 &   0.24424 &   1.79143 &   0.13920 &   0.98041 &   0.84803 &   0.09108 &   6.19365 &   0.52154 &   0.06451 \\
DFT-MD 7.24~(equal-$Z$) &  -599.07 &    35.48 &   -2.579 &     0.214 &   0.59004 &   0.19512 &   0.24498 &   1.79695 &   0.13984 &   0.98531 &   0.85239 &   0.09159 &   6.22719 &   0.52474 &   0.06492 \\
DFT-MD 7.20            &  -596.88 &    35.24 &   -2.556 &     0.211 &   0.59000 &   0.19440 &   0.24404 &   1.78994 &   0.13902 &   0.97903 &   0.84678 &   0.09093 &   6.18392 &   0.52058 &   0.06438 \\
DFT-MD 7.15            &  -596.31 &    35.18 &   -2.549 &     0.211 &   0.58999 &   0.19422 &   0.24381 &   1.78811 &   0.13881 &   0.97733 &   0.84526 &   0.09074 &   6.17202 &   0.51941 &   0.06423 \\
DFT-MD 7.15~($J_4$)     &  -587.14 &    34.18 &   -2.451 &     0.201 &   0.58985 &   0.19118 &   0.23989 &   1.75874 &   0.13537 &   0.95088 &   0.82162 &   0.08794 &   5.98975 &   0.50178 &   0.06195 \\
{\bf DFT-MD 7.13}        & {\bf -596.05} &  {\bf   35.15} &  {\bf  -2.546} &    {\bf
0.210} &  {\bf  0.58999} &   {\bf 0.19413} &   {\bf 0.24370} &   {\bf 1.78728} &
{\bf 0.13871} &   {\bf 0.97655} &   {\bf 0.84456} &   {\bf 0.09066} &   {\bf 6.16658}
&   {\bf 0.51887} &   {\bf 0.06416} \\
DFT-MD 7.13 (low-$Z$)   &  -601.72 &    35.77 &   -2.608 &     0.217 &   0.59009 &   0.19599 &   0.24610 &   1.80546 &   0.14083 &   0.99296 &   0.85924 &   0.09239 &   6.28019 &   0.52985 &   0.06558 \\
DFT-MD 7.08        &  -595.48 &    35.08 &   -2.539 &     0.210 &   0.58998 &   0.19395 &   0.24346 &   1.78542 &   0.13848 &   0.97482 &   0.84301 &   0.09047 &   6.15442 &   0.51767 &   0.06400 \\
SC 7.15           &  -589.10 &    34.86 &   -2.556 &     0.214 &   0.58993 &   0.19112 &   0.24002 &   1.76641 &   0.13699 &   0.96568 &   0.83567 &   0.09024 &   6.12279 &   0.51832 &   0.06449 \\
SC 7.15 ($J_4$)         &  -587.14 &    34.65 &   -2.534 &     0.212 &   0.58991 &   0.19048 &   0.23918 &   1.76013 &   0.13625 &   0.95997 &   0.83054 &   0.08963 &   6.08299 &   0.51443 &   0.06398 \\
\enddata
\label{tab:model_harmonics}
\tablenotetext{}{ All Love numbers for a tidal response with $q_{\rm tid}$ and $R/a$
corresponding to Jupiter's Satellite Io. Preferred interior model shown in bold face.}
\tablerefs{a. JUP230 orbit solution \cite{jacobson2003} }
\end{deluxetable}

\clearpage

\floattable
\begin{deluxetable}{l|rrrrr}
    \tablecaption{ Tidal Response for Various Satellites and Non-rotating Model}

\tablecolumns{6}
\tablewidth{0pc}
\tablehead{
\colhead{} & \colhead{\bf Io} & \colhead{Io\tablenotemark{a}} &
    \colhead{Europa} & \colhead{Ganymede} \\
    \colhead{}  &  \colhead{} &
    \colhead{non-} & \colhead{} & \colhead{} \\
    \colhead{}  &  \colhead{} & \colhead{rotating} & 
}
\startdata
\tableline
$k_{22}$  &  {\bf  0.58999}  &  0.53725  &  0.58964  &  0.58949  \\
$k_{31}$  &  {\bf  0.1941}   &  0.2283   &  0.1938   &  0.1937   \\
$k_{33}$  &  {\bf  0.2437}   &  0.2283   &  0.2435   &  0.2435   \\
$k_{42}$  &  {\bf  1.787}    &  0.1311   &  4.357    &  12.41    \\
$k_{44}$  &  {\bf  0.1387}   &  0.1311   &  0.1386   &  0.1386   \\
$k_{51}$  &  {\bf  0.9766}   &  0.0860   &  2.373    &  6.7486   \\
$k_{53}$  &  {\bf  0.8446}   &  0.0860   &  2.0289   &  5.740    \\
$k_{55}$  &  {\bf  0.0907}   &  0.0860   &  0.0906   &  0.0906   \\
$k_{62}$  &  {\bf  6.167}    &  0.0610   &  37.04    &  302.1    \\
$k_{64}$  &  {\bf  0.5189}   &  0.0610   &  1.237    &  3.487    \\
$k_{66}$  &  {\bf  0.0642}   &  0.0610   &  0.0641   &  0.0641   \\
\enddata
\label{tab:satellite_harmonics}
\tablenotetext{}{Tidal response of preferred interior model `DFT\_MD 7.13' with
    $q_{\rm tid}$ and $R/a$ for three large satellites, and for a `non-rotating'
    model with $q_{\rm rot}=0$. In bold face is the same preferred model as in
    \ref{tab:model_harmonics}. }
\tablenotetext{a}{Non-rotating model has identical density structure to rotating
model.  }
\end{deluxetable}



\end{document}